\documentclass[acmsmall,screen]{acmart}
\acmBooktitle{Proceedings of the 32nd ACM Symposium on the Foundations of Software Engineering (FSE '24), November 15--19, 2024, Porto de Galinhas, Brazil}

\usepackage{listings}
\usepackage{hyperref}
\usepackage{graphicx}
\usepackage{subcaption}
\usepackage[skins]{tcolorbox}
\usepackage{soul, xcolor}
\usepackage{color}
\usepackage{xspace}
\usepackage{multirow}
\usepackage{xurl}
\usepackage{url}
\usepackage{enumitem}
\usepackage{booktabs}
\usepackage{makecell}
\usepackage{siunitx}
\usepackage{wrapfig}
\usepackage{pifont}

\definecolor{deepblue}{rgb}{0,.2,0.6}
\definecolor{deepgreen}{rgb}{0,0.5,0}
\definecolor{deepchampagne}{rgb}{0.98, 0.84, 0.65}
\definecolor{mintgreen}{rgb}{0.6, 1.0, 0.6}
\definecolor{vividviolet}{rgb}{0.62, 0.0, 1.0}
\definecolor{mangotango}{rgb}{1.0, 0.51, 0.26}
\definecolor{dkgreen}{rgb}{0,0.5,0}
\definecolor{dkred}{rgb}{0.5,0,0}
\definecolor{gray}{rgb}{0.5,0.5,0.5}

\lstdefinestyle{cstyle} {
language=C,
basicstyle=\ttfamily\bfseries\footnotesize,
  morekeywords={virtualinvoke},
  keywordstyle=\color{blue},
  ndkeywordstyle=\color{red},
  commentstyle=\color{dkred},
  stringstyle=\color{dkgreen},
  numbers=left,
  breaklines=true,
  numberstyle=\ttfamily\footnotesize\color{gray},
  stepnumber=1,
  numbersep=10pt,
  backgroundcolor=\color{white},
  tabsize=4,
  showspaces=false,
  showstringspaces=false,
  xleftmargin=.23in,
  escapeinside={(*@}{@*)},
}
\colorlet{shadecolor}{gray!40}
\sethlcolor{shadecolor}

\newcommand{\code}[1]{\texttt{\hl{#1}}}
\newcommand{\greencheck}[0]{\textcolor{green}{\ding{52}}}
\newcommand{\redx}[0]{\textcolor{red}{\ding{55}}}
\newcommand{\graycircle}[0]{\textcolor{gray}{\ding{108}}}
\newcommand{\yellowcircle}[0]{\textcolor{yellow}{\ding{108}}}

\setcopyright{none}
\renewcommand\footnotetextcopyrightpermission[1]{}
\newcommand{\brandon}[1]{}
\newcommand{\yoshi}[1]{}
\newcommand{\joey}[1]{}

\newcommand{\tool}{\textit{VERT}\xspace}
\newcommand{\rwasm}{\texttt{rWasm}\xspace}
\newcommand{\bolero}{\texttt{bolero}\xspace}

\definecolor{dawnblue}{rgb}{0.84, 0.92, 1.0}

\definecolor{LightGray}{gray}{0.9}

\begin{document}
\title{VERT: Verified Equivalent Rust Transpilation with Large Language Models as Few-Shot Learners}

\author{Aidan Z.H. Yang}
\email{aidan@cmu.edu}
\affiliation{
\institution{Carnegie Mellon University}
\country{USA}
}
\authornote{Equal contribution.}
\authornote{Work done at Amazon.}

\author{Yoshiki Takashima}
\email{ytakashi@andrew.cmu.edu}
\affiliation{
\institution{Carnegie Mellon University}
\country{USA}
}
\authornotemark[1]
\authornotemark[2]

\author{Brandon Paulsen}
\email{bpaulse@amazon.com}
\affiliation{
\institution{Amazon Web Services}
\country{USA}
}
\author{Josiah Dodds}
\email{jldodds@amazon.com}
\affiliation{
\institution{Amazon Web Services}
\country{USA}
}
\author{Daniel Kroening}
\email{dkr@amazon.co.uk}
\affiliation{
\institution{Amazon Web Services}
\country{USA}
}

\begin{abstract}
  Rust is a programming language that combines memory safety and
  low-level control, providing C-like performance while guaranteeing the absence of undefined behaviors by default. Rust's growing popularity has prompted research on safe and correct transpiling of existing code-bases to Rust. Existing work falls into two categories: rule-based and large
  language model (LLM)-based. While rule-based approaches can
  theoretically produce \textit{correct} transpilations that maintain input-output equivalence to the original, they often yield
  unreadable Rust code that uses unsafe subsets of
  the Rust language. On the other hand, while LLM-based approaches typically produce more readable, maintainable, and safe code, they do not provide any guarantees about correctness. In this work, we present \tool, a tool that can produce readable Rust transpilations with formal guarantees of correctness. \tool{}'s only requirement is that there is Web
  Assembly compiler for the source language, which is true for most major languages. \tool{} first uses the Web Assembly compiler to obtain an \textit{oracle} Rust program. In parallel, \tool{} uses an LLM to generate a readable candidate Rust program. This candidate is verified against the oracle, and if verification fails, we regenerate a new candidate transpilation until verification succeeds. We evaluate \tool by transpiling a suite of 1,394 programs taken from competitive programming style benchmarks. Our results show that \tool{} significantly improves an LLM's ability to generate correct Rust transpilations. Combining Anthropic's Claude-2 and \tool increases Rust transpilations passing property-based testing from 31\% to 54\% and bounded model-checking from 1\% to 42\%
  compared to using Claude alone. In addition, we evaluate \tool{}'s ability to generate non-trivial safe Rust on programs taken from real-world C projects that make significant use of pointers. Our results provide insights into the limitations of LLMs to write safe Rust.

\end{abstract}


\maketitle
\section{Introduction} 
\begin{figure*} \centering
\includegraphics[width=\textwidth]{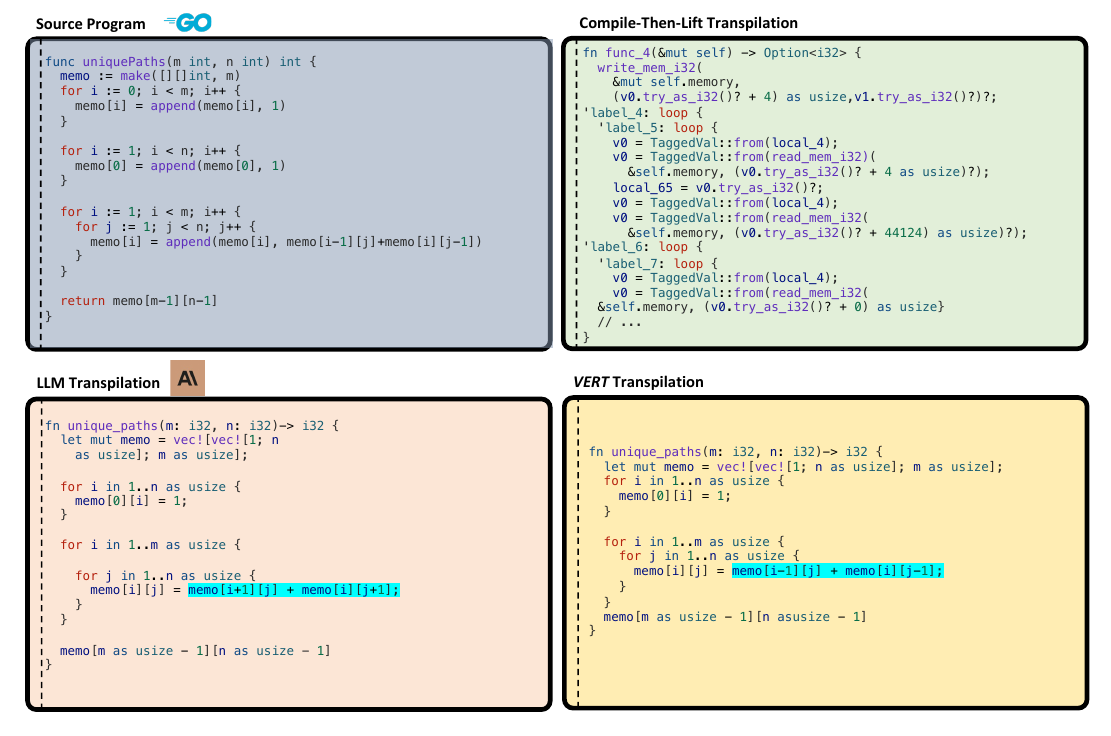}
\vspace*{-3mm}
\caption{\small Examples of transpilations from a source Go program using compile-then-lift, Anthropic's Claude LLM, and~\tool{}.}
\label{fig:flow}
\end{figure*}

\brandon{It may be confusing to readers that we claim "Verified Equivalent" in the title, but then in the abstract we talk about property-based checking and bounded checking, neither of which are verified in the absolute sense}
Rust is a memory- and type-safe programming language that has
performance on par with low-level languages like C.
It is often referred to as a ``safer C'' because the Rust type checker can guarantee
the absence of undefined behavior.
Microsoft estimates that 70\% of all their security bugs are
due to memory-safety issues~\cite{microsoft-cve-study}, which could be
mostly or entirely eliminated if the code were written in Rust.
Citing Rust's security benefits, Rust has been used in major open source projects
such as Firecracker~\cite{firecracker},
and Linus Torvalds recently announced Rust will be a supported
language for Linux kernel development~\cite{rust-linux}.

Rust's security and performance benefits have fueled interest in
automatically transpiling existing code written in other languages
into Rust~\cite{C2Rust, szafraniec2022code}. Existing works on transpilation broadly fit into two categories: \textit{rule-based} and \textit{large language model (LLM)-based}. Rule-based approaches use hand-written rules and
algorithms that translate a target program into a new language,
typically in a statement-by-statement fashion. Rule-based approaches
have human-understandable implementations that could, in theory, be proved correct. 
However, as we will show, they often result in
unreadable and unidiomatic code that does not take full advantage of the target language's useful features, such as efficient native types.

On the other hand, LLM-based approaches train an LLM that takes a program in one language as input and attempt to output an equivalent program in the target
language~\cite{roziere2020unsupervised}. LLM-based approaches tend to produce code that is similar to their training data, and thus, if the model is trained on high quality, human written, code, the model will usually produce high quality, idiomatic transpilations~\cite{yang2024revisiting}. However, these approaches come with no formal guarantees that the resulting code will maintain input-output equivalence with the original~\cite{consistency, xu2024hallucination, pan2024lost}.

In this work, we focus on \textit{general}, \textit{verified}, and \textit{readable} transpilation to Rust.  By \textit{general}, we mean that it can apply to most major languages. By \textit{verified}, we mean that input-output equivalence of the final transpilation has been verified against the source program in some way. The verification techniques we experiment with are property-based testing (PBT), bounded model checking, and unbounded model checking. By \textit{readable}, we mean that the transpilation resembles human-written code. While readability is a highly subjective measure, we show examples in Figure~\ref{fig:flow} that we believe makes this claim self-evident.

To the best of our knowledge, the only \textit{general} rule-based approach for transpiling to Rust that could \textit{theoretically} guarantee equivalence is to \textit{compile-then-lift}. In this approach, we first compile the source program to an intermediate language like LLVM or Web Assembly, and then lift the intermediate language to the target language (Rust in our case). For example, Web Assembly compilers exist for most major languages (e.g. C, C++, Rust, Java, Go, C\#, PHP, Python, TypeScript, Zig, and Kotlin), and the recent work \rwasm~\cite{bosamiya2022provably} can lift Web Assembly to Rust. While this approach is very general and, at least in theory, can guarantee equivalence, compile-then-lift approaches generally can only produce code that is as readable as the intermediate language itself, which for LLVM and Web Assembly is virtually unreadable~\cite{romano2020wasim, garba2019saturn}. We give an example transpilation using a Web Assembly compiler and \rwasm in Figure~\ref{fig:flow}. As can been seen, \rwasm produces Rust that looks like assembly rather than a high-level language. 

There are also many works on transpiling without equivalence guarantees~\cite{soar2020, szafraniec2022code, roziere2020unsupervised, weisz2021perfection, weisz2022better, ni2021soar}, mainly using language models. While these approaches are also general and usually produce readable code, language models are notorious for outputting subtly incorrect code~\cite{xu2024hallucination}. We show an example language model transpilation in Figure~\ref{fig:flow} using Anthropic's Claude-2, which is a state-of-the-art general purpose LLM. We can see the transpilation is far more readable than the result of the compile-then-lift approach. However, the LLM has changed a \textsc{-} to a \textsc{+}, which may escape human review. Such subtle errors may be difficult to debug and only manifest in corner cases.

To overcome these limitations, we combine rule-based and LLM-based transpilation with formal verification tools, and implement our approach in a tool~\tool{}.
Our algorithm takes a source program as input, and outputs a transpilation that is verified equivalent relative to a rule-based transpilation. Notably, \tool{} does not require any additional input beyond the source program. The main assumption of \tool{} is that the language of the source program has a Web Assembly compiler.

\tool{} first creates an \textit{oracle} Rust transpilation by using the source language's Web Assembly compiler and \rwasm, as previously described. This transpilation is equivalent by construction, but is unreadable. Next, we leverage an LLM to produce a candidate final transpilation, which is far more readable, but may have implementation errors, ranging from syntax errors to subtle logic errors. We then enter an iterative repair and verify process. We first attempt to fix compilation errors by applying a combination of hand-crafted rules and re-prompting the LLM to re-transpile the source program until the program compiles. Once compiled, we attempt to verify equivalence using one of the previously mentioned verification techniques. If verification succeeds, then we stop and output the program. However, if verification fails, which is usually the case, we re-prompt the LLM to transpile the program.

We evaluate \tool{} on 1,394 transpilation tasks with source languages in C++, C, and Go curated from prior work~\cite{szafraniec2022code, zhang2023ownership}.
We focus on C++ and C since these two languages are often used for similar tasks as Rust~\cite{von2024integrating, jung2020understanding}. We further evaluate on Go as Rust is often the transpilation target when cross-platform support is a hard requirement~\cite{jain2023choose}.
We experiment with three state-of-the-art LLMs as the underlying LLM for \tool{}, namely CodeLlama-2~\cite{codellama}, StarCoder~\cite{li2023starcoder}, and Anthropic Claude-2~\cite{claude}. 
With Claude-2 as the underlying LLM, our results show that \tool{} can produce transpilations that pass bounded verification for 42\% of these programs, and differential testing for 54\% of these programs. Moreover, \tool{} improves the capabilities of existing LLMs -- \tool{} is able to produce a verified transpilation 45\% more often on average (43\% for C++, 46\% for C, and 43\% for Go) compared to using an LLM alone.

To measure \tool{}'s ability to produce non-trivial \textit{safe} Rust, we gather and manually label an additional 14 programs that make significant use of pointers from the benchmarks of prior work on C to Rust transpilation~\cite{C2Rust, zhang2023ownership, emre2021translating}. Our results on these additional benchmarks show that \tool{} can produce Rust with relatively simple ownership models for small (less than 36 LoC) programs. 


In summary, our main contributions are as follows.
\begin{itemize}
[leftmargin=*]
\item{\textbf{\tool.} We propose and implement an end-to-end technique that can transpile any language that compiles to Wasm into human-readable Rust. Our data and tool are available for open-source.\footnote{\url{https://zenodo.org/records/10927704}}}

\item{\textbf{Verified Equivalence with Input Code} We use Wasm
generated from original input as a reference model and perform
equivalence checking by automatically injecting test harnesses, allowing the user to verify that the LLM
translation is free of hallucinations.}

\item{\textbf{Empirical evaluation}. We evaluated \tool on a set of real world programs and competitive programming solutions, which include 569 C++ programs, 520 C programs, and 305 Go programs. We perform an extensive evaluation of several LLMs directly (CodeLlama-2), with fine-tuning (StarCoder), and with instruction-tuned few-shot learning (Anthropic Claude-2) on different source code languages. 
}
\end{itemize}

\section{Background}

We give a brief introduction of the key aspects of our tool. In particular: Rust, the \rwasm compilation strategy, and auto-regressive large language models.

\subsection{Rust}
%
Rust is a systems programming language with a focus on performance,
reliability, and safety.  Rust's main goal is to eliminate memory
safety errors through a \textit{memory-ownership} mechanism.
%
Rust's memory-ownership mechanism associates each value in memory with
a unique \textit{owner} variable, which guarantees safe static memory collection.
In particular, when we want to create a variable that aliases a value
(i.e., creating a new pointer to the value), we must transfer ownership
to the new variable, either temporarily or permanently.
Rust programs are challenging to write and synthesize as these ownership rules
must be obeyed for the program to even compile. LLM-based Rust
synthesis has the additional challenge that these typing rules are not found in popular languages that make up majority of the training
dataset.




\subsection{Migrating to Rust}
Given the memory-safety properties of Rust, there is a strong
incentive to migrate existing codebases to Rust. While several notable
projects have been rewritten in Rust~\cite{zhang2023review}, the
translation to Rust remains a challenge owing to the enormous manual effort. For C
to Rust translation in particular, several tools have been developed
to automatically translate C functions to Rust~\cite{C2Rust,
  zhang2023ownership, emre2021translating}. These tools use the
semantic similarity between C and Rust and apply re-writing rules to
generate Rust code. However, these re-write rules are specific to the
source language, and do not scale to multiple languages, especially
those whose semantics is not similar to Rust. To the best of our
knowledge, no rule-based automatic translator exists from a
garbage-collected language like Go or Java to Rust.

In contrast to transpilers, \rwasm differs significantly in its
intent. \rwasm converts Web Assembly (Wasm) programs into Rust and
leverages the memory-safety properties of safe Rust as a sandbox to
eliminate the Wasm runtime overhead. Since many programming
languages already target Wasm as an intermediate
representation~\cite{akinyemi2023awesome}, we can leverage \rwasm for multi-language support.

\subsection{Rust Testing and Verification}
To establish trust in the LLM-output, we perform equivalence
verification of two Rust programs. We use existing tools that operate on Rust to
prove equivalence between the LLM-generated and the \rwasm-generated
oracle Rust programs. We use \bolero, a Rust testing and verification
framework that can check properties using both Property-Based Testing
(PBT)~\cite{fink1997property} and Bounded Model
Checking~\cite{clarke2001bounded,DBLP:conf/tacas/ClarkeKL04}. An example of a \bolero harness for
checking equivalence between an LLM-generated and a trusted reference program is given in Fig.~\ref{code:bolero-eq}.

\begin{figure}[h]
\centering
 \begin{lstlisting}[numbersep=5pt,xleftmargin=21pt,numberstyle=\scriptsize,basicstyle=\footnotesize\ttfamily,firstnumber=1]
#[test]
#[cfg_attr(kani, kani::proof)]
fn eq_check()  {
    bolero::check!()
        .with_type()
        .cloned()
        .for_each(|(a, b) : (i32, i32)| llm_fn(a, b) == reference_fn(a, b));
}
\end{lstlisting}
\caption{\small An example bolero harness checking equivalence between
  2 functions.}
\label{code:bolero-eq}
\end{figure}

Using a harness like the one in Fig.~\ref{code:bolero-eq}, \bolero can check
properties via two different methods. The first method is by random
PBT. PBT works by randomly generating inputs to the function under
test, running the harness with these inputs, and asserting the desired
properties (e.g. equivalence between two functions). PBT repeatedly runs this procedure to
check the property over the range of inputs. PBT is a valuable tool for catching implementation bugs,
however it is generally infeasible to run PBT for long enough to exhaust all possible inputs to a program.

\bolero performs model checking through Kani~\cite{vanhattum2022kani},
a model-checker for Rust. When run with Kani, \bolero produces
symbolic inputs rather than random concrete inputs. Executing the
harness with symbolic inputs, we can cover the entire space of inputs
in one run and the model-checker ensures the property holds for all
possible inputs. Since symbolic execution does not know how many times
loops are run, Kani symbolically executes loops up to an user-provided
bound. To prove soundness of this bound, Kani uses {\it unwind checks}
asserting that loop iteration beyond the bound is not reachable. We
say that a verification is {\it bounded} if unwind checks are turned
off and exhaustiveness is not known. Conversely, a {\it full}
verification includes unwind checks and thus is exhaustive with
respect to all reachable executions. Even with full verification, the
harness may not be able to cover enough values of unbounded types like
vectors.

While Kani can prove properties of programs, complex programs can take
too long to prove.  Conversely, PBT runs exactly as quickly as the
program runs, but is not exhaustive. Given the complementary
properties of PBT and Kani, we allow users to use both tools, using
\bolero and marking the harness for both PBT ({\code{\#[test]}}) and
model-checking ({\code{kani::proof}).

\subsection{Large Language Models}

Deep learning (DL) has recently shown promise for program generation~\cite{vaithilingam2022expectation, nijkamp2022codegen}. The DL models that can achieve the closest capabilities to human-written results are large language models (LLMs), such as GPT-4~\cite{openai2023gpt4}.
LLMs train billions of parameters using a massive amount of training data.
LLMs' effectiveness for code generation~\cite{chen2021evaluating} suggests that LLMs are capable of performing specialized software engineering tasks, such as program language transpilation. 

Most modern LLMs are attention-based models. Attention-based models use the Transformer architecture \cite{vaswani2017attention}. In a Transformer architecture model, tokens exchange information across all other tokens using an attention matrix. LLMs typically produce text in a left-to-right manner (auto-regressive model), producing each token given its prefix context. In a auto-regressive models, the learned attention matrix is partially masked out. Specifically, the generation of a new token depends only on its prefix context (i.e., tokens on the left), and its suffix context (i.e., tokens on the right) are hidden (i.e., masked out). After a model finishes training all trainable parameters in a user specified time, a model can make predictions on future tokens on a sequence of tokens the model has never seen before. 




\section{Methodology}
\label{section:methodology}

\begin{figure*}
\centering
\includegraphics[width=\textwidth]{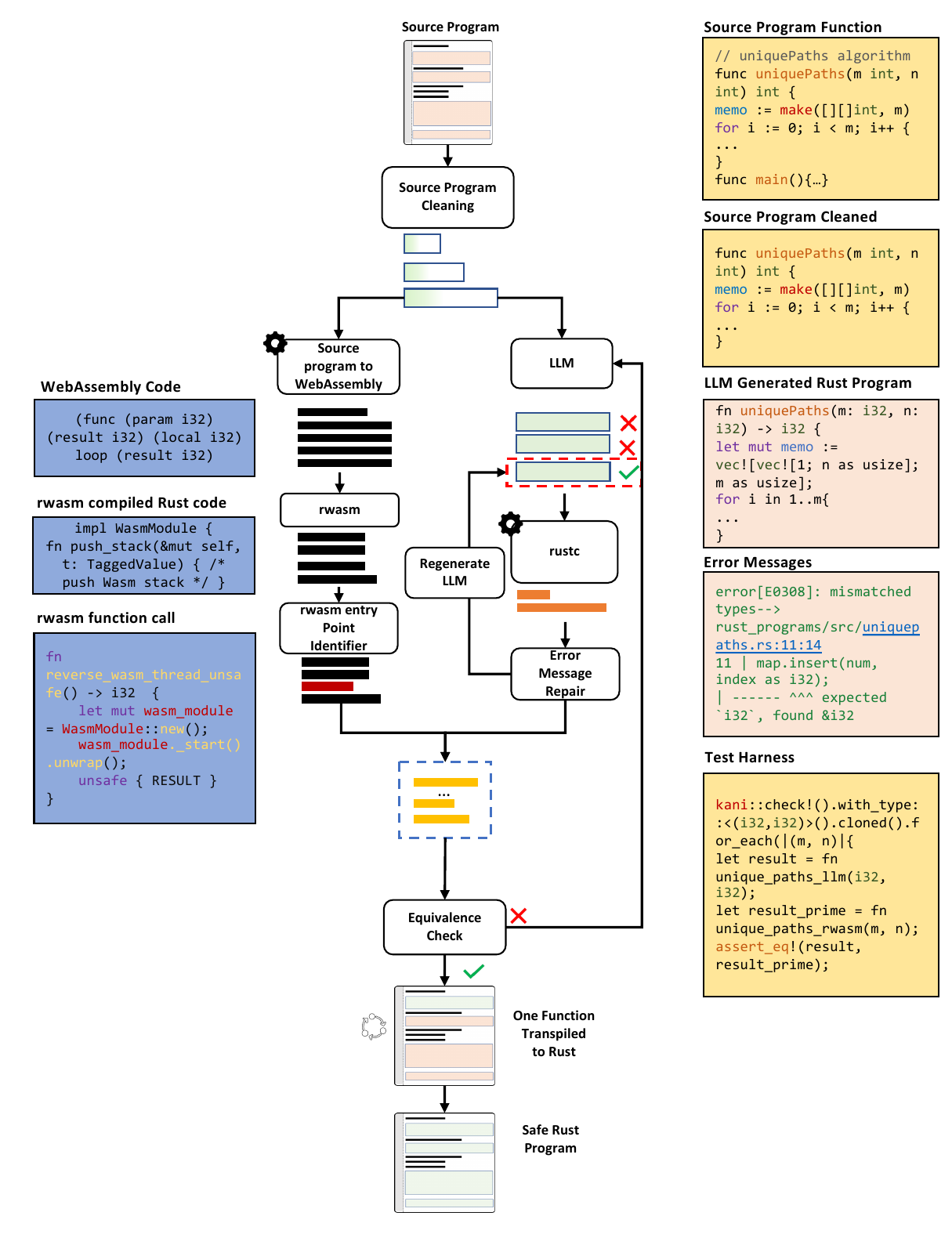}
\vspace*{-3mm}
\caption{\small \tool’s architecture, which takes as input a source program and produces a formally equivalent Rust program}
\label{fig:overview}
\end{figure*}

In this section, we describe the key ideas behind our universal transpilation technique.
Figure~\ref{fig:overview} gives an overview of \tool's entire pipeline. The technique takes as input a source program, and outputs a verified equivalent Rust program. As shown in Figure \ref{fig:overview}, we parse the program into separate functions during the cleaning phase, then split the pipeline into two paths. The first path outputs an LLM-generated Rust transpilation. The second path produces \rwasm Rust code that is compiled directly from the original program through Wasm. Finally, we create a test harness based on the original program to verify equivalence of the two paths' outputs, and only after a successful verification we output a correct and maintainable Rust transpilation. In the following sections, we describe each component of \tool in further detail.

\subsection{Program repair on LLM output}
\label{sec:repair}
LLMs often produce incorrect code. When prompting an LLM for Rust code, any slight mistake could cause the strict Rust compiler ($\bold{rustc}$) to fail. Fortunately, $\bold{rustc}$ produces detailed error messages when compilation fails to guide the user to fix their program. We create an automatic repair system based on $\bold{rustc}$ error messages. For each error, we first classify the error into one of three main categories: syntax, typing, and domain specific. 

As seen in Figure \ref{code:error1}, an example syntax error generated by the LLM is the wrong closing delimiter. For $\bold{rustc}$ to successfully compile, all syntax errors must be resolved. We track the error code location (e.g., line 10 in Figure \ref{code:error1}), and we use the $\bold{rustc}$ provided initial delimiter to guide our repair strategy. For this case, we know to use the right curly bracket \code{\}} to replace \code{]} on line 10. 

Typing error messages in Rust generally have a similar structure. In particular, error messages are usually of the form \code{expected type a, found b}. Figure~\ref{code:error2} shows the LLM generate a pass-by-reference variable \code{\&i32} while $\bold{rustc}$ expects a pass-by-value \code{i32}. Using the compiler message's error localization and suggestion line (characterized by the keyword \code{help:}), we replace the variable \code{num} by \code{\&num}. 

Finally, domain-specific errors are compilation errors that are specific to the program. The error messages for domain-specific errors do not share the same structure, and therefore we only use the $\bold{rustc}$ error message suggestion line to generate a repair. In Figure~\ref{code:error3}, which shows the error message for an immutable assignment, the suggestion line indicates that if the variable \code{x} is converted to a mutable object, the immutable assignment error would be solved. Using this suggestion line, we replace \code{x} by \code{mut x}, and observe that the program compiles. It is often the case that even with the error message suggestion line, we cannot generate a repair that fixes all errors. In these cases, we regenerate an LLM output using the error as part of the new prompt and restart the process. Our error-guided repair is significantly faster than the LLM generation (discussed in Section~\ref{section:results}), so we only regenerate an LLM output after exhausting all $\bold{rustc}$ helper messages.

\begin{figure}[t]
\centering
    \begin{lstlisting}[numbersep=5pt,xleftmargin=21pt,numberstyle=\scriptsize,basicstyle=\footnotesize\ttfamily,firstnumber=1, language=C]

For more information about this error, try 'rustc --explain E0433'.
error: mismatched closing delimiter: ']'
  --> roman_to_integer_test.rs:1:33
   |
1  | fn roman_to_int(s: &str) -> i32 {
   |                                 ^ unclosed delimiter
...
10 |     ]);
   |     ^ mismatched closing delimiter
\end{lstlisting}
\vspace*{-2mm}
\caption{\small Syntax error}
\label{code:error1}
\end{figure}

\begin{figure}[t]
\centering
\begin{lstlisting}[numbersep=5pt,xleftmargin=21pt,numberstyle=\scriptsize,basicstyle=\footnotesize\ttfamily,firstnumber=1, language=C, language=C]
error[E0308]: mismatched types-->
rust_programs/src/uniquepaths.rs:11:14
11 | map.insert(num,index as i32);| ------ ^^^ expected '&i32', found i32
                                               help: consider borrowing here: '&num'
\end{lstlisting}
\vspace*{-2mm}
\caption{\small Mismatched type error}
\label{code:error2}
\end{figure}

\begin{figure}[t]
\centering
\begin{lstlisting}[numbersep=5pt,xleftmargin=21pt,numberstyle=\scriptsize,basicstyle=\footnotesize\ttfamily,firstnumber=1, language=C]
error[E0384]: cannot assign to immutable argument 'x'
  --> reverse_integer_test.rs:16:3
   |
1  | fn reverse(x: i32) -> i32 {
   |- help: consider making this binding mutable: 'mut x'
...
16 |         x = x / 10;
   |         ^^^^^^^^^^ cannot assign to immutable argument

error: aborting due to 2 previous errors
For more information about this error, try 'rustc --explain E0384'.
\end{lstlisting}
\vspace*{-2mm}
\caption{\small Immutable assignment error}
\label{code:error3}
\end{figure}

\begin{figure}
\centering
\begin{lstlisting}[numbersep=5pt,xleftmargin=21pt,numberstyle=\scriptsize,basicstyle=\footnotesize\ttfamily,firstnumber=1, language=C]
func callReverse() int {
  result := reverse(123)

  if result == 321 {
    return 0
  } else {
    return 1
  }
}
\end{lstlisting}
\vspace*{-2mm}
\caption{\small An entry point for the {\tt reverse} function}
\label{code:gotestreverse}
\end{figure}

\subsection{Transpilation Oracle Generation}
Since the LLM output cannot be trusted on its own, we create an
alternate trusted transpilation pipeline for generating a reference Rust
program against which the LLM output is checked. The alternate
pipeline does not need to produce maintainable code, but it needs to translate the source language into Rust using a reliably correct rule-based
method. We use Wasm as the intermediate representation because
many languages have compilers to Wasm, allowing it to serve as the
common representation in the rule-based translation. Once the
input programs are compiled to Wasm, we use
\rwasm~\cite{bosamiya2022provably}, a tool that translates from Wasm to
Rust by embedding the Wasm semantics in Rust source code. While the
original authors intended \rwasm as a sandboxing tool that leverages
the memory safety properties of safe Rust, we use it to generate
trusted Rust code with same semantics as the original input.

\subsection{Mutation Guided Entry Point Identification}
\brandon{The subsection title here sounds like we are talking about an approach for testing equivalence, but it is really about how we identify the entry point in the \rwasm output}
\brandon{Is it assumed that the user provides the program as shown in Figure~\ref{code:gotestreverse}? Or do we automatically generate it?}
Given the assembly-like output of \rwasm, we must perform analysis to
identify the entry point of the \rwasm transpiled function. \tool
provides the option for the user to manually identify the entrypoint,
but we can find it automatically using a simple heuristic, such as a function call or a single test case. We note that this heuristic could be generated automatically using LLMs or search-based software testing and thus we can assume an entrypoint generator in the source language. \tool uses a function call in the source language with constant
inputs to the function to be transpiled and an assertion on the output
of that function. One such function is given in Fig.~\ref{code:gotestreverse}.

\begin{figure}
\centering
\begin{lstlisting}[numbersep=5pt,xleftmargin=21pt,numberstyle=\scriptsize,basicstyle=\footnotesize\ttfamily,firstnumber=1, language=C]
fn func_4(&mut self, ) -> Option<i32> {
  // ...
  let mut local_3 : i32 = 0i32;
  let mut local_4 : i32 = 0i32;
  v0 = TaggedVal::from(321i32);
  // mutant: v0 = TaggedVal::from(654i32);
  local_3 = v0.try_as_i32()?;
  v0 = TaggedVal::from(123i32);
  // mutant: v0 = TaggedVal::from(456i32);
  local_4 = v0.try_as_i32()?;
  // ...
}
\end{lstlisting}
\vspace*{-2mm}
\caption{\small The difference between the original \rwasm output and the
  mutated one (highlighted).}
\label{code:gotestwasm}
\end{figure}

We leverage this function call to identify the input and output of the
function. While one option for such analysis is to perform
decompilation, we find that a mutation-guided approach is sufficient
for our purposes. In Fig.~\ref{code:gotestreverse}, we know that the
input is \code{123} and the output is \code{321}. Now, we wish to
identify the equivalent constant in the \rwasm output. While it is
possible to just perform a linear scan of the \rwasm output for this
constant, that risks spurious matches, especially for simple types
like \code{i32}. Instead, we guide this identification by leveraging
the function call and mutating it. Suppose we swap \code{123} with
\code{456} and \code{321} with \code{654} and re-transpile with
\rwasm. These constants will change, but the rest of the \rwasm output
remains the same. Taking the diff, we can identify inputs and outputs
by what changed. The diff in the \rwasm output is shown in
Fig.~\ref{code:gotestwasm}.

\subsection{Equivalence Harness Generation}

In our final step, we generate harnesses to check for equivalence
given the input and output locations. We define equivalence here in
functional terms: for all inputs, running both functions yields no
crashes and identical outputs. To check this property holds, we automatically
generate a wrapper to the Wasm function and a harness where the
LLM-synthesized and wrapped \rwasm functions are called with the same
inputs, and the outputs are asserted to be equal. To ensure this
equivalence holds for all inputs, we leverage property-based testing
with random inputs and model-checking with symbolic inputs. For the
remainder of this section, we refer to both of them together as ``the
input.''

The wrapper consists of two parts: input injection, and output
checking. Since the original program runs with a set of constant
inputs, we must replace these constant inputs with the inputs of the
harness like \code{input: i32}. The challenge here is to inject the
inputs into the middle of the Rust code representing a Wasm
module. Instead of replacing the parameters to the function, we use
globals in Rust to inject the inputs right at the location where
constants used to be. An example is given in
Fig.~\ref{code:eqharness}, with \code{func\_4} being the Wasm
equivalent of the test. \tool replaces constant inputs with global
reads, generalizing the test and allowing us to vary the inputs fed
into the Wasm-generated function. Note that, while this injection
requires unsafe code, it is fine as this is only done in the
oracle and the oracle is discarded once the equivalence is checked.

Now that we can feed various inputs into the Wasm function, we must
also provide a way to assert that the output is equal. Recall that in
Fig.~\ref{code:eqharness}, the output is compared to a baseline and
$0$ is returned if the check succeeds. Because of this comparison
check, it is sufficient to inject the baseline value and then leverage
the check to assert that the return value is $0$. Injection is done in
the same way as the inputs.

We note that while this approach is sound, it may falsely identify
some equivalent programs as faulty due to semantic differences between
Rust and the target language, or between the target language and
\rwasm{} embedding. We note two cases where we permit the analyst to
add assumptions. First, when the input type is an unsigned integer. In
this case, we have a mismatch were Wasm has only signed integers. So
the output of \rwasm{} will represent unsigned integers by encoding it
in signed. However, the true valid range will be smaller (\code{u32}
will become \code{i64} to the full range of \code{u32} values but the
extra bits are not used). In this case, it is soundly permissible to
assume that the values lie in the valid range of \code{u32}. Another
case of valid assumptions occurs with strings: strings in C are ASCII
while in Rust are Unicode. Therefore, a valid range of Rust strings
will crash a C-derived Wasm module spuriously. We assume the string's
range to valid ASCII only.

\begin{figure}
\centering
\begin{lstlisting}[numbersep=5pt,xleftmargin=21pt,numberstyle=\scriptsize,basicstyle=\footnotesize\ttfamily,firstnumber=1, language=C]
static mut INPUT_1 = 0;
static mut OUTPUT_1 = 0;
impl WasmModule {
  /// returns 0 if the output matches
  fn func_4(&mut self, ) -> Option<i32> {
    // ...
    let mut local_3 : i32 = 0i32;
    let mut local_4 : i32 = 0i32;
    v0 = TaggedVal::from(unsafe {INPUT_1});
    local_3 = v0.try_as_i32()?;
    v0 = TaggedVal::from(unsafe {OUTPUT_1});
    local_4 = v0.try_as_i32()?;
    // ...
  }
}
/// equivalence-checking harness.
fn equvalence() {
  bolero::check!()
  .for_each(|(input: i32)| {
    let llm_fn_output = llm_generated_reverse();
    unsafe {
      INPUT_1 = input;
      OUTPUT_1 =  llm_fn_output;
    }
    let mut wasm_module = WasmModule::new();
    wasm_module._start().unwrap();
    assert!(wasm_module.func_4().unwrap() == 0);
  });
}
\end{lstlisting}
\caption{\small Equivalence-checking harness for {\tt func\_4}.}
\label{code:eqharness}
\vspace*{-5mm}
\end{figure}

\subsection{Equivalence Checking}
With the equivalence checking harness built, we must now drive the
harness and check that the equivalence property holds for all
inputs. \tool provides 3 equivalence checking techniques with
increasing levels of confidence and compute cost. This procedure is
shown in Fig.~\ref{fig:eval}. First, we run the equivalence-checking
harness with PBT using \bolero up to the time limit, generating random
inputs and checking equivalence of the outputs. If the candidate
diverges from the oracle, then PBT will return the diverging input as
a counterexample. If no counterexample is found within the time limit,
we say this candidate passes PBT.

If the PBT stage succeeds, we now perform bounded verification with
Kani. In the bounded verification phase, we run Kani with an unwinding bound of $k$ and \textit{no unwind checks}. This means that paths up
to $k$ loop iterations is exhaustively explored, but any divergences
between the candidate and the oracle with traces containing more than
$k$ loop iterations are missed. We run this phase for 120 seconds, and
terminate with 3 potential results. First, Kani returns with a
counterexample that causes the oracle and the candidate to diverge or
one of the two to crash. Second, Kani does not return with an answer
within the time limit, which we also consider to be a failure as we
cannot establish bounded equivalence. Finally, Kani verifies that,
limited to executions with at most $k$ loop iteration, there are no
divergences or crashes. We consider the third case alone to be
successful. Bounded verification does not check whether $k$ is
exhaustive.

If bounded verification succeeds, we perform full verification. \tool
increases the unwinding bound until verification can be achieved with
all checks enabled, including unwind checks that ensure the unwinding
is fully exhaustive and the program cannot go beyond the code that was
unwound. This ensures the equivalence and safety properties holds for
every input the harness generates. Once again, three outcomes are
possible. First, Kani can return a counterexample. Second, Kani can
fail to return an answer within the time limit. Finally, Kani
successfully verifies the harness. Again, we consider the third case
alone to be successful. However, unlike with bounded verification,
successful full verification guarantees that the translation is
without any error. If the oracle crashes at any point in the
equivalence checking, \tool provides the user with a counterexample
which can be used to diagnose the crash in the original program.

We support complex types through their primitive parts. Given a struct
or enum, that Kani or PBT does not initially support, we construct
values of that type by abstracting the primitive parameters of that
type and any required discriminants for enums. For types of finite
size, this is sufficient. However, we provide bounded support for
handling vector types. The challenge here is to vary the length of the
vector in the \rwasm output, which is done by having a fixed-length
vector of varying inputs and then pruning the length down to the
actual length dynamically. Our approach is sound and complete for
primitive types, and by extension, any type that comprises solely of
primitive types such as tuples of primitives. For unbounded types like
vectors, hashmaps and user-defined types containing such, \tool
synthesizes harnesses that generate inputs up to the size encountered
in the sample function call. As a limitation, any divergences that
require bigger vector than encountered will be missed.

\subsection{Few-shot Learning}
\label{sec:fewshot}
The main focus of this work is on verifying the output of LLMs for program transpilation, and not LLM prompt engineering. Therefore, we keep the prompts simple and short. Complicated and repeated querying of the same prompts do not provide additional benefits on the accuracy of outputs for small sized models, and too expensive for an average practitioner for industry sized models (i.e., Anthropic Claude).
To achieve few-shot learning on our transpilation queries, each failed transpilation attempt provides its equivalence checking counter examples as a few-shot learning example for future transpilation attempts. 

Figure.~\ref{code:prompt} shows our template for few shot learning. We start with querying the LLM to refactor the source code into safe Rust. Although we filter for safe Rust LLM output, we experimentally found that asking the LLM to always produce safe Rust gives more accurate results. We prompt the LLM to use the same argument and return types as the original, and can compile without external dependencies. Finally, we collect the counter examples from prior failed equivalence checks as part of the prompt. Specifically, we ask the LLM to consider the specific inputs that caused a test or verification failure from the previous iterations. We observed that providing specific inputs as information to the LLM results in subtle bug fixes within the program output.

\begin{figure}
\centering
\begin{lstlisting}[numbersep=5pt,xleftmargin=21pt,numberstyle=\scriptsize,basicstyle=\footnotesize\ttfamily,firstnumber=1, language=C]
{Original code}
...
Safe Rust refactoring of above code in {language}, with code only, no comments. 
Use the same function name, same argument and return types.
Make sure the output program can compile on its own.
// If there exists counter examples from prior failed equivalence checking
Test that outputs from inputs {counter_examples} are equivalent to source program.
\end{lstlisting}
\caption{\small LLM Prompt template.}
\label{code:prompt}
\vspace*{-5mm}
\end{figure}

\section{Evaluation}
\label{section:eval}

\begin{figure*} \centering
\includegraphics[width=\textwidth]{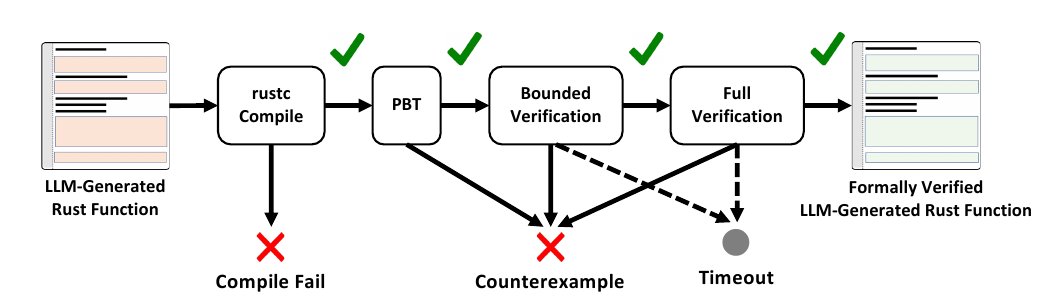}
\vspace*{-3mm}
\caption{\small Evaluation procedure.}
\label{fig:eval}
\vspace*{-5mm}
\end{figure*}

In this section, we present our approach and results for the following research questions.

\noindent\textbf{RQ1. How does \tool perform vs. using the respective LLM by itself?}
We evaluate our technique's performance on a benchmark dataset,
showing that \tool significantly increases the number of verified equivalent transpilations vs. using the LLM by itself.

\noindent\textbf{RQ2. How does each component of \tool's approach impact its performance?}
We conduct an ablation analysis, which shows that our prompting and
error-guided refinement helps produce more well-typed and more correct
programs.
We further measure the runtime performance of each part of \tool, showing that
time costs of error-guided refinement is reasonable and \tool
spends most of the time in verification. 

\noindent\textbf{RQ3. Does \tool produce safe, readable, and idiomatic Rust transpilations?}
To evaluate \tool's ability to produce safe Rust, we collect programs from real world C projects that make use of pointers. In addition, we report on the frequency of linter warnings of transpilations produced by \tool, and compare lines of code between the translations produced by \tool, \rwasm, and CROWN~\cite{zhang2023ownership}, a rule-based C to Rust transpiler. \tool's transpilations do not produce linter warnings, and has far fewer lines of code than the other approaches.

\noindent\textbf{RQ4. How extensible is \tool{} to future verifiers?}
While we use Kani~\cite{vanhattum2022kani} for most of the benchmarks to leverage automation, we encountered a large number of timeouts. We show that \tool{} is able to work with multiple verifiers by using Verus~\cite{verus} and show
that manual verification is possible albeit costly.

\subsection{Setup}

\subsubsection{LLMs}

We use the following LLMs to generate the candidate transpilations in \tool{}:
\begin{itemize}[leftmargin=*]
\item{\textbf{TransCoder-IR~\cite{szafraniec2022code}}: A language
    model trained by low-level compiler intermediate
    representations (IR) for the specific purpose of programming
    language translation. TransCoder-IR improves upon the TransCoder
    model ~\cite{roziere2020unsupervised} by incorporating IR into the
    training data and decompiling into IR as a training target. Both
    TransCoder and TransCoder-IR are trained on roughly 2.8 million repositories from GitHub\footnote{https://console.cloud.google.com/marketplace/details/github/github-repos}. Since TransCoder-IR's input is the original
    code alone and no prompt is taken, we do not perform error-guided few-shot prompting.
    To the best of our knowledge, TransCoder-IR is the only LLM-based general transpilation tool for Rust. Therefore, we use TransCoder-IR as baseline for our evaluation.
    }

\item \textbf{CodeLlama-2~\cite{codellama}}: A 13B parameter
  model initialized from Llama-2~\cite{touvron2023llama}, then further fine-tuned on 500 billion tokens of code data.

\item \textbf{StarCoder Fine-tuned~\cite{li2023starcoder}}: A 15.5B parameter
  model trained on 1 trillion tokens sourced from The
  Stack~\cite{kocetkov2022stack}. StarCoder prompted achieves the
  highest HumanEval~\cite{chen2021evaluating} score of 40.8 over
  comparable open-source LLMs, such as
  LLaMA-65B~\cite{touvron2023llama} with a score 23.7 and
  CodeGen-16B~\cite{nijkamp2022codegen} with a score of 29.3. To investigate the effectiveness of Rust fine-tuning on prior LLMs, we fine-tune StarCoder for transpilation using LeetCode problems that have solutions for Rust, C, C++, and Go. In total, we collect solutions in each language for 94 LeetCode problems. We fine-tune the LLM to take C, C++, or Go solution as the input, and produce the corresponding Rust solution as output.

\item \textbf{Anthropic Claude-2~\cite{claude}}: A production-grade, proprietary LLM accessible through Anthropic's APIs with roughly 130 billion parameters. Claude-2 costs about \$0.0465 per
  thousand tokens.
\end{itemize}

\subsubsection{LLM Fine-tuning}
The availability of Rust code in open source is scarce as compared code written in most other programming languages. Incoder~\cite{fried2022incoder} estimates that Rust is only the 20th most common language in their training database, which is a 159GB code corpus taken from Github BigQuery \footnote{https://cloud.google.com/blog/topics/public-datasets/github-on-bigquery-analyze-all-the-open-source-code}. Due to the lack of Rust data available on open-source, we opt to not train an LLM targeted at Rust code generation. Instead, we directly use an off-the-shelf industry grade LLM, and also fine-tune on a separate open-source pretrained LLM. Specifically, we use Anthropic Claude-2 \footnote{https://www.anthropic.com/product} for the industry grade LLM, and StarCoder~\cite{li2023starcoder} for the pretrained LLM. 

We use light weight and parameter efficient adapter layers~\cite{hu2023llm, yang2024large, peft} for fine-tuning StarCoder. Instead of retraining StarCoder entirely, we taken the final hidden states of StarCoder and add adapter layers at the end using small amounts of data.
We collect 94 LeetCode type question solutions in C, C++, Go and Rust. Although there are existing code bases for all four languages, we find that LeetCode has the most consistent translation between other languages and Rust. We were able to collect 94 LeetCode questions of which have a direct translation between all 3 languages.
For each LeetCode type question, we have a corresponding source program (written in Go, C, or C++), and a target program (written in Rust). We encode all code words into tokens using the GPT-2 tokenizer. We fine-tune with 4 Transformer layers, 300 total epochs, and a final model dimension of 6144.

\subsubsection{Benchmark selection}
We draw our benchmarks from two sources. Our first source is the benchmark set from TransCoder-IR \cite{szafraniec2022code}, which is primarily made up of competitive program solutions. In total, this benchmark set contains 852 C++ programs, 698 C programs, and 343 Go programs. We choose this dataset to avoid potential data-leakage (i.e., LLM memorization)~\cite{biderman2024emergent} in our evaluation. We note that the Rust programs produced by TransCoder-IR were released after June 2022, which is the training data cutoff date of our chosen LLMs~\cite{li2023starcoder, claude, codellama}. 
We select programs from the TransCoder-IR dataset that can directly compile to Wasm using \rwasm. After filtering, we collect a benchmark set of 569 C++ programs, 506 C programs, and 341 Go programs. 
These types of benchmarks are common for evaluating LLMs' coding ability. However, the programs themselves often do not make extensive use of pointers, so they do not adequately challenge \tool{}'s ability to generate safe Rust.

To provide insight into~\tool{}'s ability to write safe rust, we gather 14 additional pointer-manipulating C programs from prior work on C to Rust transpilation~\cite{C2Rust, zhang2023ownership, emre2021translating}. We note, however, that the benchmarks in these prior works use open-source programs written before our chosen LLM's training data cutoff (June 2022). To avoid LLM data-leakage, we select and customize snippets from these C projects to transpile to Rust. We manually label the input output pairs for each snippet for verifying equivalence on the transpiled Rust programs. Many of the benchmarks we select involve multiple functions. The explicit goal when selecting benchmarks from these projects is to discover the limitations of~\tool{} in terms of writing safe Rust, therefore we gather benchmarks of increasing complexity in terms of the number of pointer variables, and the number of functions in the benchmark. We present several complexity metrics for the benchmarks and discuss them in more detail in Section~\ref{section:results}. 

In total, we evaluate our approach on \textbf{569 C++} programs, \textbf{520 C} programs, and \textbf{341 Go} programs.

\subsubsection{Evaluation Metrics}
Neural machine translation (NMT) approaches use
metrics that measure token similarity between the expected output and the actual output produced by the LLM, in which a higher score indicates the two outputs have many tokens in common. While these approaches are often meaningful when applied to natural language, for programming languages, small differences in the expected output and actual output could result in different compilation or run-time behavior.
Conversely, two programs that share very few tokens (and hence have a very low text similarity score) could have identical compilation or run-time behavior.

For programming languages, metrics based off of passing tests have been proposed. Roziere et al.~\cite{roziere2020unsupervised} and Szafraniec
et al.~\cite{szafraniec2022code} use the computational accuracy (CA)
metric, which counts a translation as correct if it passes a series of
unit tests. However, there is no accepted standard for the
number of required passing tests when using the CA metric. Furthermore, the CA metric does not take into
account the quality or coverage of the unit tests.

To improve upon the metrics used for prior NMT approaches and remove the overhead of writing high-coverage unit tests, we use formal methods to measure the correctness of the output. We insert the LLM-generated code and \rwasm-generated code in an equivalence-checking harness that asserts equal inputs lead to equal outputs. An example of such a harness is given in Figure~\ref{code:eqharness}. Our full procedure is shown in
Fig.~\ref{fig:eval}. Since the three metrics used are significantly
slower than checking a series of unit tests, we set a time limit for
our metrics. For all three metrics, we set a 120 seconds limit. For
PBT, no counterexamples within 120 seconds counts as success. For
Bounded and Full verification, success requires establishing verified equivalence within 120 seconds. If any of the three verification step fails, \tool terminates.

\subsubsection{Environment}

The experiments for all benchmarks were run on an Ubuntu 22 instance with 32 Intel Xeon 2.30 GHz processors, 240GB RAM, and 4 Tesla V100 GPUs.

\subsection{Results}
We present results on the TransCoder-IR benchmarks in Table~\ref{table:results}. We present \tool{} operating in three different modes. \textit{Single-shot} means that \tool{} uses the LLM \textit{once} to create a single candidate transpilation, and then proceeds directly to verification. If verification fails, then \tool{} does not attempt to regenerate. 
\textit{Few-shot repair} means that, if verification fails, then \tool{} will prompt the LLM to regenerate the transpilation repeatedly. In each iteration, we apply the syntactic repair described in Section~\ref{sec:repair} to the output of the LLM. Finally, \textit{Few-shot repair \& counter examples} means that we use counter examples produced by previous failed verification attempts as part of the LLM's few-shot learning, as described in Section~\ref{sec:fewshot}. \textit{Few-shot repair \& counter examples} only works for instruction-tuned models. We re-prompt the LLM up to 20 times for few-shot modes. For each LLM and each mode of \tool{}, we report the number of transpilations that compiled and that passed the various verification modes. As seen in Table~\ref{table:results}, we only perform \textit{single-shot} for Transcoder-IR (baseline) to replicate results from prior work. We perform \textit{few-shot repair} on CodeLlama2 and StarCoder fine-tuned to investigate the effectiveness of few-shot and rule-based repair on open-source, non-instruction tuned LLMs. Finally, we perform \textit{single-shot, few-shot repair, and few-shot repair with counter examples} with Anthropic Claude-2 to investigate how each part of \tool impacts an instruction-tuned LLM's ability to perform Rust transpilation.

\label{section:results}

\noindent\textbf{RQ1. How does \tool perform vs. using the respective LLM by itself?}

\begin{table*}[t]
\caption{\small \tool performance across with different LLMs and modes.}
\begin{tabular}{p{2.1cm}|p{1.2cm}p{2.5cm}rrrr}
\toprule
 \textbf{LLM} &  \textbf{Source Lang} & \textbf{Technique} & \textbf{Compiled} & \textbf{PBT} & \textbf{Bounded-ver.} & \textbf{Full-ver.}\\
\midrule

\multirow{3}{*}{\makecell{Transcoder \\ IR \\ (Baseline)}}
& \hspace*{-5mm}\makecell{C++ \\ (569)}
& Single-shot             & 107  & 23 & 3 & 0 \\
\cmidrule{2-7}
& \hspace*{-5mm}\makecell{C \\ (520)}
& Single-shot             & 101  & 14 & 1 & 0 \\
\cmidrule{2-7}
& \hspace*{-5mm}\makecell{Go \\ (341)} 
& Single-shot             & 24  & 3 & 0 & 0 \\
\midrule

\multirow{3}{*}{\makecell{CodeLlama2 \\ 13B}} 
& \hspace*{-5mm}\makecell{C++ \\ (569)}  
& Few-shot repair   & 307  & 25 & 6 & 2 \\
\cmidrule{2-7}
& \hspace*{-5mm}\makecell{C \\ (520)} 
& Few-shot repair   & 160 & 18 & 4 & 2 \\
\cmidrule{2-7}
& \hspace*{-5mm}\makecell{Go \\ (341)}
& Few-shot repair    & 104 & 15 & 2 & 0 \\
\midrule

\multirow{3}{*}{\makecell{StarCoder \\ fine-tuned \\ 15.5B}}
& \hspace*{-5mm}\makecell{C++ \\ (569)}
& Few-shot repair             & 253  & 79 & 8 & 2 \\
\cmidrule{2-7}
& \hspace*{-5mm}\makecell{C \\ (520)}
& Few-shot repair    & 179 & 76 & 4 & 2 \\
\cmidrule{2-7}
& \hspace*{-5mm}\makecell{Go \\ (341)} 
& Few-shot repair    & 134 & 59 & 2 & 0 \\
\midrule
\multirow{9}{*}{\makecell{Anthropic \\ Claude-2 \\ 130B}}
& \multirow{4}{*}{\makecell{C++ \\ (569)}}
& Single-shot            & 240  & 55 & 6 & 0 \\
& & Few-shot repair    & 539 & 292 & 41 & 2 \\
& & Few-shot repair \& counter examples (VERT) & \textbf{539} & \textbf{295} & \textbf{233} & \textbf{19} \\
\cmidrule{2-7}

& \multirow{4}{*}{\makecell{C \\ (520)}}
& Single-shot            & 239  & 49 & 6 & 0 \\
& & Few-shot repair    & 339 & 195 & 29 & 4 \\
& & Few-shot repair \& counter examples (VERT) & \textbf{339} & \textbf{209} & \textbf{193} & \textbf{15} \\
\cmidrule{2-7}

& \multirow{4}{*}{\makecell{Go \\ (341)} }
& Single-shot             & 126  & 26 & 3 & 0 \\
& & Few-shot repair     & 276 & 157 & 39 & 4 \\
& & Few-shot repair \& counter examples (VERT) & \textbf{317} & \textbf{195} & \textbf{159} & \textbf{9} \\
\bottomrule
\end{tabular}
\label{table:results}
\end{table*}

\begin{table*}[t]
\caption{\small \tool's average runtime per component for a Single-program translation}
\begin{tabular}{llr}
\toprule

\textbf{Component type} & \textbf{Component} & \textbf{Time (s)}\\
\midrule\multirow{3}{*}{LLM}
& Transcoder-IR & 8 \\
& CodeLlama-2 & 43 \\
& Starcoder fine-tuned & 45 \\
& Anthropic Claude & 30 \\

\midrule\multirow{3}{*}{Rust compilation}
& $\texttt{rustc}$ & <1 \\
& Error guided repair & 1 \\
& $\texttt{rwasm}$ & <1 \\

\midrule\multirow{3}{*}{Testing and verification}
& PBT & 25 \\
& Bounded-ver. & 52 \\
& Full-ver. & \textbf{67} \\

\bottomrule
\end{tabular}
\label{table:time}
\end{table*}

As seen in table~\ref{table:results}, \tool with Claude-2 compiles for 76\% more programs for C++, 75\% for C, and 82\% for Go as compared to baseline (i.e., Transcoder-IR).
\tool with Claude-2 can pass PBT for 49\% more programs for C++, 37\% for C, and 56\% for Go as compared to baseline.
\tool with Claude-2 can pass bounded verification for 40\% more programs for C++, 37\% for C, and 47\% for Go as compared to baseline.
For passing full verification, \tool with Claude-2 can transpile 19 C++ programs, 15 C programs, and 9 Go programs. Transcoder-IR cannot pass full verification on any of the tested programs. \tool with both CodeLlama2 and StarCoder fine-tuned also improve upon baseline on number of programs passing compilation, PBT, bounded verification, and full verification. We observe that few-shot learning with rule-based repair on general code-based LLMs can perform more accurate Rust transpilations than an LLM trained with transpilation as its main target.

To confirm that \tool yields a statistically significant improvement over baseline, we perform a Wilcoxon rank test~\cite{woolson2007wilcoxon}, which indicates if the metric performance between \tool and baseline are statistically different. We use the Wilcoxon signed-rank test to see if the statistically significant difference is also positive (i.e., our approach is different and better as measured by our three metrics). We observe Wilcoxon signed-rank p-values ranging from $\num{1e-5}$ to $\num{4e-5}$ for PBT, bounded-verification, and full-verification.

\begin{tcolorbox}
[colback=white,colframe=black,arc=0pt,boxrule=0.5pt,title=RQ1 Summary,boxsep=2pt,left=1pt,right=1pt,top=1pt,bottom=1pt,fonttitle=\bfseries]
\tool with CodeLlama2, StarCoder fine-tuned, and Anthropic Claude-2 can produce more PBT, bounded verification, and full verification passing Rust transpilations than baseline. In particular, \tool with Claude-2 can pass bounded verification for 40\% more programs for C++, 37\% for C, and 47\% for Go as compared to baseline.
\end{tcolorbox}

\noindent\textbf{RQ2. How does each component of \tool impact its performance?}
Table~\ref{table:results} shows the transpilation results across CodeLlama-2 and StarCoder fine-tuned in a few-shot setting.
We observe that \tool with CodeLlama-2 and StarCoder fine-tuned improve over Transcoder slightly for compilable Rust translations. Since Rust is an underrepresented language in all LLMs trained on GitHub open-source repositories and The Stack dataset~\cite{kocetkov2022stack}, we see that light-weight fine-tuning on a small dataset shows immediate improvement. In particular, we observe that StarCoder fine-tuned has fewer transpilations than CodeLlama-2 passing compilation, but more transpilations than CodeLlama-2 passing bounded verification. Fine-tuning with Rust code has an immediate impact on transpilation accuracy. 
StarCoder's results are limited by its ability to pass compilation, even with \tool's \texttt{rustc} error guided program repair in place. \tool with StarCoder fine-tuned compiles 47\% fewer programs for C++, 41\% fewer for C, and 63\% fewer programs for Go as compared to \tool with Claude-2. While adding fine-tuning on Rust syntax increases the number of compilable translation generated, we observe that an industry-grade LLM with more trainable parameters and a larger training dataset performs significantly better for our metrics.

We observe that \tool using few-shot plus repair with either StarCoder fine-tuned or Claude-2 yields better transpilation across all our three languages and three metrics. In particular, few-shot plus repair with Claude-2 passes 43\% more PBT checks for C++, 46\% more for C, and 43\% more for Go as compared to single-shot with Claude-2. Table~\ref{table:results} does not show single-shot results for CodeLlama-2 and StarCoder fine-tuned as we observed no transpilations passing PBT. Few-shot plus repair with Claude-2 passes 6\% more bounded-verification checks for C++, 4\% more for C, and 12\% more for Go as compared to single-shot with Claude-2. We find that the few-shot prompting for Claude-2 yields a greater improvement over single-shot compared to our repair technique. For C++ and C in particular, few-shot and repair with Claude-2 does not provide any additional passes on bounded verification nor full verification as compared to only few-shot with Claude-2. We observe that few-shot learning with counter examples of failed previous verification attempts provides the largest improvements on both bounded-verification and full-verification. Modern LLMs that are instruction-tuned can learn to generate more correct program when given specific test failures in few-shot settings.

Table ~\ref{table:time} shows the average runtime of each of \tool's
components across our entire evaluation dataset. We observe that in
the non-timeout failure cases (i.e., Kani does not establish
equivalence within 120s), Kani's full verification (full-ver.) uses an
average of 67 seconds per program. Kani's bounded verification uses an
average of 52 seconds per program, and Bolero's property testing uses
an average of 25 seconds per program. Of the LLMs, both CodeLlama-2 and StarCoder use about 3 seconds per each prompt attempt, and Anthropic Claude-2 about 2
seconds. Not counting the failure cases (i.e., the LLM does not
generate any program that can pass equivalence after 20 attempts), we
observe an average of 15 tries before the LLM can achieve
compilation. Transcoder-IR uses 8 seconds on average per
transpilation, which we prompt only one time as the baseline of our
evaluation.

\begin{tcolorbox}
[colback=white,colframe=black,arc=0pt,boxrule=0.5pt,title=RQ2 Summary,boxsep=2pt,left=1pt,right=1pt,top=1pt,bottom=1pt,fonttitle=\bfseries]
Our ablation study shows that fine-tuning an LLM with Rust yields a higher accuracy of transpiled programs, as seen by a higher number of programs passing PBT and bounded verification by StarCoder fine-tuned compared to CodeLlama2.
However, few-shot learning with counter examples provides the largest improvements on transpilation accuracy. Finally, we observe that \tool spends most of its runtime in verification.
\end{tcolorbox}

\noindent\textbf{RQ3. Does \tool produce safe, readable, and idiomatic Rust transpilations?}
To measure \tool{}'s ability to generate safe Rust, we use \tool{} few-shot + repair with Claude-2 to transpile the 14 C programs described previously. Table~\ref{table:results_pointer} presents the results of \tool{} on these 14 benchmarks as well as several metrics that provide a rough idea of the complexity of the benchmarks. Specifically, we present the number of pointer variables, function definitions, LoC, and the number of structs defined in each benchmark. The \code{avl\_*} benchmarks are taken from a library that implements an AVL tree. The \code{brotli\_*} benchmarks from the Brotli compression library. The \code{buffer\_*} benchmarks allocate and resize a buffer respectively. The \code{ht\_*} benchmarks compute a hash key, and create a hash table, respectively, the \code{libcsv\_*} benchmarks initialize a struct with pointer variables, and retrieve members from the struct. \code{libtree} determines if an array of pointers to \code{int64}s is sorted. \code{urlparser} parses a url. 


\tool can produce transpilations for 7 of the 14 C programs that pass
PBT, and 2 of those can pass bounded verification. Two benchmarks
cannot pass compilation due to Rust's borrow checker
(\texttt{ht\_create} and \texttt{urlparser)}. In particular, \tool was
unable to generate safe Rust on \texttt{ht\_create} due to
transferring a variable into byte representation in two lines of
code. 

The results show that \tool{} tends to struggle as the programs get
larger, have more pointer variables, and also on programs with
multiple functions. Still on smaller programs, the LLM can still
determine basic ownership needs. For example, it can determine if a
variable or parameter reference needs a \code{mut} permission. On the
\code{avl\_insert} benchmark, the LLM successfully assigns ownership
to the newly created node. 
To evaluate the readability, we compare lines of code in the transpilations produced by \tool, \rwasm, and CROWN~\cite{zhang2023ownership}, the rule-based C to Rust transpiler.
After running \code{rustfmt}~\cite{rustfmt}, the official formatter of Rust, CROWN's output is more than 5x larger than \tool, and \rwasm's output more than 10x as large.
Given the strong negative
association between LoC and code readability~\cite{buse_learning_2010}, we conclude that
\tool's outputs are more readable than CROWN and \rwasm.

To evaluate the idiomaticity, we run Clippy\footnote{https://doc.rust-lang.org/clippy/}, the official linter of Rust, on \tool's transpilations. Clippy checks Rust code for potential issues in the categories of correctness (e.g. checking if an unsigned int is greater than 0), performance (e.g. unnecessarily using a Box or collection type), stylistic (e.g. not following naming conventions, unnecessary borrowing), and conciseness (e.g. unnecessary type casting or parentheses). On average, Clippy produces 10.9 warnings per function for CROWN, and 372 warnings per function for \rwasm. Clippy does not produce any warnings on \tool's transpilations, thus we conclude that they are reasonably idiomatic. 

\tool targets the broader and more difficult problem of general,
verified translation to Rust, whereas CROWN only targets unsafe to
safe rust (after running C2Rust~\cite{C2Rust}) without
verification. For the 14 programs, both \tool and CROWN output safe
Rust. However, \tool’s output is more Rust-native than CROWN’s, using
standard Rust types while CROWN and C2Rust use C-foreign
types/functions. \tool’s lack of reliance on C-foreign functions is a
qualitative strength. \tool’s output is more self-contained and
reviewable to Rust programmers~\cite{astrauskas2020programmers}. \tool
can catch buggy C API calls in the input program instead of
translating the incorrect API calls to Rust $libc::$ calls that remain
buggy. Finally, we note that CROWN assumes correctness on their
output, and only runs deterministic test suites on 6 example
benchmarks (corresponding to 4/14 of our selected pointer
benchmarks). \tool performs three layers of equivalence checking on all
its output.

\begin{table*}[t]
\caption{\small Benchmark information and results on the 14 C pointer manipulation programs. The symbols \greencheck{}, \redx{}, and \graycircle{} indicate pass, fail (with counterexample), and timeout.}
\begin{tabular}{lrrrrrrr}
\toprule
\textbf{Benchmark} & \textbf{Structs}  & \textbf{Functions} & \textbf{\makecell{Pointer \\ variables}}& \textbf{LOC}   & \textbf{Compiled} & \textbf{PBT}&\textbf{\makecell{Bounded\\ Ver.}}\\
\midrule
avl\_minvalue & 1 & 1 & 4& 17 & \greencheck{}& \greencheck{}&\greencheck{}\\
avl\_insert & 1 & 2 & 4& 30 & \greencheck{}& \greencheck{}&\graycircle{}\\
avl\_rotate & 1& 3& 7& 32 & \greencheck{}& \greencheck{}&\graycircle{}\\
avl\_delete & 1 & 4 & 27& 111 & \redx{} & \redx{}&\redx{}\\
brotli\_parse\_int  & 0 & 1 & 2 & 15 & \greencheck{}& \greencheck{}&\graycircle{}\\
brotli\_filesize  & 1 & 1 & 1 & 28   & \greencheck{} & \redx{}&\redx{}\\
buffer\_new & 1 & 1 & 3& 16 & \greencheck{}& \greencheck{}&\graycircle{}\\
buffer\_resize & 1 & 3 & 3& 22 &\greencheck{}& \redx{}&\redx{}\\
ht\_hashkey & 0 & 1 & 2& 13 & \greencheck{}& \greencheck{}&\greencheck{}\\
ht\_create & 2& 1& 3& 36 & \redx{} & \redx{}&\redx{}\\
libcsv\_get\_opts & 1 & 1 & 1& 29 &\greencheck{}& \redx{}&\redx{}\\
libcsv\_init & 1& 1& 4& 55 &\redx{} & \redx{}&\redx{}\\
libtree & 0 & 1& 1& 7 & \greencheck{}& \greencheck{}&\graycircle{}\\
urlparser & 0& 9 & 28& 158 &\redx{}& \redx{}&\redx{}\\

\bottomrule
\end{tabular}
\label{table:results_pointer}
\end{table*}

\begin{tcolorbox}
[colback=white,colframe=black,arc=0pt,boxrule=0.5pt,title=RQ3 Summary,boxsep=2pt,left=1pt,right=1pt,top=1pt,bottom=1pt,fonttitle=\bfseries]
\tool can produce transpilations for 7 of the 14 C programs that pass PBT, and 2 of those can pass bounded verification. \tool tends to struggle as programs have more pointer variables, or have multiple functions. However, \tool is far more readable than prior work. \tool produces 5X less LoC than CROWN, 10x less LoC than \rwasm, and its transpiled Rust programs do not show any linter warnings. 
\end{tcolorbox}

\noindent\textbf{RQ4. How extensible is \tool{} to future verifiers?}
We observe in Table~\ref{table:results} that few transpiled Rust
programs can pass full-verification with Kani, which is a bounded
model checker (BMC). Full-verification using a BMC results in complete
unrolling of a program, which does not scale to programs that loop
over their inputs.  We consider using Verus~\cite{verus} as the
verifier instead of Kani. Given that the verification failures are due
to Kani unrolling loops, we use Verus to specify loop invariants and
algebraic specifications for proving equivalence.

\tool{} handles multiple verifiers for the equivalence checking
step. This is useful when different verifiers have different strengths
and weaknesses. For example, Kani is a bounded model checker so loops
are difficult to verify. Verus is a autoactive
verifier~\cite{verus}, so it can verify loops more
effectively via invariants, but at the cost of lower automation. To
understand the need for an autoactive verifier, we run lightweight
analysis using regex matching on our benchmarks that Kani failed to
verify. 86\% of the benchmarks had explicit iteration over an
input. Furthermore, this is an undercount because it ignores loops
that happen in API calls and library functions like \code{strcpy}.

\begin{table}[h]
  \caption{\small We manually verify 5 timeout cases in RQ4 using
    Verus.  LoC is the lines of code while Spec. Loc is the lines of
    specification. No Spec. LoC is given when verification is not
    successful. The symbols \greencheck{} and \yellowcircle{} indicate
    pass and feature limitation respectively. }
\begin{tabular}{lrrr}
\toprule
\textbf{Benchmark} & \textbf{Code LoC} & \textbf{Spec. LoC}  & \textbf{\makecell{Verus Ver.}}\\
\midrule
avl\_insert & 30 & - &\yellowcircle{}\\
avl\_rotate & 32 & - &\yellowcircle{}\\
brotli\_parse\_int & 15 & 92 &\greencheck{} \\
buffer\_new & 16 & 1 &\greencheck{}\\
libtree & 7 & 6 &\greencheck{}\\
\bottomrule
\end{tabular}
\label{table:results_verus}
\end{table}

Given the manual effort required to manually verify each of the LLM's
outputs, we limit ourselves to the 5 cases in the CROWN benchmark where PBT
succeeded but bounded Kani failed. The results are show in the
Table~\ref{table:results_verus}.
We succeed in verifying equivalence for 3 of the 5 failed benchmarks
while we fail to verify AVL benchmarks due to Verus's limitations on
returning mutable pointers. In each of the successful cases, checking
equivalence required fully specifying and verifying the resulting
programs. Given that, the spec to code ratio varied substantially
depending on the benchmark. For \code{brotli\_new} the program simply
allocates a large buffer that caused Kani to time out. So the
specification is a one line, and would have probably passed with Kani
using the new contract feature or with a lower buffer
size. \code{libtree} benchmark is a function that, given a list of
integers, checks if it is sorted. This one required a quantifier-based
specification to assert an invariant over the range of the array that
has checked as sorted. While the spec size is not substantial, the
presence of quantifiers over unbounded arrays will have been difficult
to specify with Kani. The final and heaviest benchmark is
\code{brotli\_parseint}, which required 92 lines of specification and
supporting proof. This function parses an integer from an array of
chars, and specifying that required recursive specs involving
exponentiation that would be difficult with Kani. The sheer size of
the spec also stems from proofs required to show that the spec is
well-behaved while recursing over the array. Overall ratio of spec to
code is 2.6.

\begin{figure}
\centering
\begin{lstlisting}[numbersep=5pt,xleftmargin=21pt,numberstyle=\scriptsize,basicstyle=\footnotesize\ttfamily,firstnumber=1, language=C]
spec fn right_parse(s: &Vec<char>, upto: int, value: int) -> bool
    decreases upto,
{
    if 0 <= upto && upto <= s.len() {
        if upto > 0 {
            (value % 10 == (s[upto - 1] as u32 - '0' as u32)) &&
              right_parse(s, upto - 1, value / 10, )
        } else { value == 0 }
    } else {false}
}
proof fn right_parse_continues(s: &Vec<char>, upto: int, value: int)
    requires
        valid_vector(s),
        right_parse(s, upto, value),
        0 <= upto < s.len(),
    ensures
        right_parse(s, upto + 1, (value * 10) + ((s[upto] as u32) - ('0' as u32))),
{ /* Verus figures out by definition you 
can append a digit by parsing one more char. */}
\end{lstlisting}
\caption{Specification of \code{brotli\_parseint}. We prove this spec
  on both the LLM and reference programs.}
\label{code:verus-parseint-spec-parse}
\vspace*{-5mm}
\end{figure}

We go over \code{brotli\_parseint} detail to describe the supporting
specification proofs for equivalence. We give the specification used
to prove \code{brotli\_parseint} function correct in
Fig.~\ref{code:verus-parseint-spec-parse}. The main specification is
\code{right\_parse}, which defines parsing of the array from right to
left upto tthe given index. The induction proof
\code{right\_parse\_continues} extends this specification over the
array of \code{char}. We also prove the absence of arithmetic
overflows in the output program by relating the maximum array length
to the size of the parsed integer. We omit the precise specification
for conciseness.

\begin{figure*}
\centering
\begin{lstlisting}[numbersep=5pt,xleftmargin=15pt,numberstyle=\scriptsize,basicstyle=\footnotesize\ttfamily,firstnumber=1, language=C]
fn llm_parseint(s: &Vec<char>, low: u32, high: u32) -> (ret: Option<u32>)
    ensures
        valid_vector(s) && ret.is_some() ==> right_parse(s, s.len() as int, 
        ret.unwrap() as int),
        valid_vector(s) && ret.is_some() ==> low <= ret.unwrap() <= high,
        !valid_vector(s) ==> ret.is_none()
{
    let mut value = 0u32;
    let mut i: usize = 0;
    while i < 5
        invariant
            valid_vector(s) && value < exp(i as int) &&
            
            right_parse(s, i as int, value as int),
    {
        let mut c: char = s[i];
        if (c as u32) < ('0' as u32) || (c as u32) > ('9' as u32) {
            return None;
        }
        assert(i < 6);
        assert(value < exp(6)) by {
            exp_monotone_any(value, i as int, 6);
        };
        assert(exp(6) == 1000000) by (compute_only);
        value = 10 * value + (c as u32 - '0' as u32);
        i += 1;
    }
    ... // Return None for error
    return Some(value);
}
\end{lstlisting}
\caption{Code of \code{brotli\_parseint}. We prove the
  \code{right\_parse} specification from
  Fig.~\ref{code:verus-parseint-spec-parse} onto the function with
  invariants. Note that both the original code in C and the LLM output
  assumes input length of at most 5.}
\label{code:verus-parseint-code}
\vspace*{-5mm}
\end{figure*}

We give the LLM's output and supporting specifications in
Fig.~\ref{code:verus-parseint-code}. We use the \code{ensures} to
specify that a valid vector parses correctly and invalid or
out-of-range values are not returned. The main loop verifies through
the invariant that the integer parsed out is correct up to the current
index of the array. For this to hold, we also trivially specify that
the vector remains valid through the loop. Finally, we put an
invariant that the integer parsed is below $10^i$, which we use to
show no integer overflow occurs. The rest of the requirements
dispatches naturally through the conditional cases. We note that full
specification, while possible, is not compatible with the automatic
nature of the tool. We expect that further improvements, such as
automatic invariant synthesis through
LLMs~\cite{yao2023leveraging} might make this approach more
amenable to users who have neither the specification nor the formal
methods expertise to use tools like Verus.

\begin{tcolorbox}
  [colback=white,colframe=black,arc=0pt,boxrule=0.5pt,title=RQ3
  Summary,boxsep=2pt,left=1pt,right=1pt,top=1pt,bottom=1pt,fonttitle=\bfseries]
We manually verify 5 programs with Verus that previously timed-out
using Kani. We succeeded in verifying equivalence for 3 of the 5
failed benchmarks by leveraging loop invariants.
\end{tcolorbox}
%

\section{Related Work}

\subsection{Language model transpilers}
Recent advances in language modelling using code as training data have shown that LLMs can perform code completion~\cite{desai2016program} and generate code based on natural language~\cite{raychev2014code} with impressive effectiveness. Large Language Models (LLMs) have raised performance on these tasks using significantly more trainable parameter and training data~\cite{chen2021evaluating}.
Transcoder and Transcoder-IR~\cite{roziere2020unsupervised, szafraniec2022code} use unsupervised machine translation to train neural transcompilers. As both Transcoder versions are trained on program translation pairing, they perform better on program translation tasks than similar sized but generic auto-regressive LLMs.

However, recent work shows that LLMs can generate buggy and vulnerable programs ~\cite{pearce2021empirical, chen2021evaluating, black2021gpt}. Transcoder-IR~\cite{szafraniec2022code} show in their evaluation that a significant portion of their translated code do not compile, especially for target programming languages that are underrepresented in their training data (e.g., Rust). Our work helps both generate memory-safe code and establish equivalence, to safely harness the code output of an LLM.

\subsection{Rust transpilers}
Prior Rust transpilers convert C/C++ to Rust.  C2Rust~\cite{C2Rust}
automatically converts large-scale C programs to Rust while preserving
C semantics. Citrus~\cite{citrus} and Bindgen~\cite{bindgen} both
generate Rust FFI bindings from C libraries, and produce Rust code
without preserving C semantics. Bosamiya et
al.~\cite{bosamiya2022provably} embedded WebAssembly (Wasm) semantics
in safe Rust code for the Rust compiler to emit safe and executable
Rust code. Bosamiya et al.~\cite{bosamiya2022provably} implemented all
stages of their tool in safe Rust, and no stage of compiler needs to
be further verified or trusted to achieve safety. Our work focuses on
generating readable and maintainable Rust code with Rust semantics
directly.

Existing tools for making unsafe Rust safer focus on converting raw pointers to safe references. Emre et al.~\cite{emre2021translating} localized unsafe Rust code from C2Rust and converted unsafe Rust to safe Rust by extracting type and borrow-checker results from the \texttt{rustc} compiler. Zhang et al.~\cite{zhang2023ownership} converts unsafe Rust from C2Rust to safe Rust by computing an ownership scheme for a given Rust program, which decides which pointers in a program are owning or non-owning at particular locations. Zhang et al.~\cite{zhang2023ownership} evaluates their tool CROWN on a benchmark suite of 20 programs, and achieve a median reduction rates for raw pointer uses of 62.1\%.
Ling et al.~\cite{ling2022rust} removed non-essential ``unsafe'' keywords in Rust function signatures and refined the scopes within unsafe block scopes to safe functions using code structure transformation. Our work is the first to propose a general Rust transpiler that does not depend on the source language's memory management properties to produce safe Rust.

\subsection{Equivalence Verification}
Equivalence verification has been studied in settings where two copies
of should-be-equivalent artifacts are available. Churchill
et al. applied equivalence checking to certify compiler optimizations
with program
alignment~\cite{churchill_semantic_2019,churchill_sound_2017}. Antonopoulos
et al. leverage an extension to Kleene Algebra with Tests (KATs) for
a more formal calculus~\cite{antonopoulos_algebra_2023}. Rule-based
approaches have been combined with machine learning by Kommrusch for
automated equivalence
verification~\cite{kommrusch_self-supervised_2023}. Conversely, Dahiya and Bansal leverage equivalence-checking in black-box settings without relying on the translator~\cite{dahiya_black-box_2017}. Our work uses equivalence verification for program language translation. Our work is the first to combine LLMs with established equivalence checking techniques (i.e., property based testing and bounded model checking) to generate both readable and verified code.


\section{Limitations and discussion}
\label{sec:threats}

Threats to \textit{internal validity} are concerned with the degree of
confidence on our dependent variables and our results. As \tool uses
the LLMs StarCoder and Claude, which take as training data GitHub
repositories up to June 2022, we can not fully mitigate the bias introduced by potential \textit{training data contamination} -- i.e. the possibility that the competitive coding questions in our evaluation datasets could be included in their training data. To mitigate this threat, we select and evaluate our tool on Rust solutions written and labeled after June 2022.
Furthermore, our evaluation is primarily to show that our iterative repair procedure can significantly improve the number of correct transpilations produced by an LLM, and that we can verify equivalence between \rwasm Rust and the LLM produced Rust. This result is not affected by training data contamination. Thus we believe potential training data contamination does not invalidate our results.

Threats to \textit{external validity} lie in whether results on our
benchmarks will generalize to real-world contexts. One of our
benchmark sets is a collection of competitive programming solutions across
three languages, used by Transcoder-IR~\cite{szafraniec2022code}. Although the collected programs cover a wide range of input and output types, real-world code bases are often much more complicated than competitive solution
programs. To reduce this threat, we further evaluate on 14 selected functions
from real world programs taken from the \textit{Crown} and \textit{Laertes} benchmark~\cite{zhang2023ownership, emre2021translating}.




\tool's equivalence checking is performed in 3 stages with increasing
guarantees: PBT, bounded, and full verification. The first 2 phases do
not always provide any exhaustive guarantees. The quality of PBT depends
heavily on the inputs generated, and thus the seed used to generate
the random inputs. Bounded verification may miss bugs or divergences
in LLM and \rwasm-generated programs if they occur beyond the loop
unwinding bound. Full verification provides exhaustive guarantees but
few programs actually reach that stage due to performance limitations
of Kani. As Kani improves, the performance of unbounded verification will improve as well.

\section{Conclusion}
Rust is a growing language with C-like performance, but provides further safety guarantees. Rust's improvements on security and performance over other languages have prompted recent research on transpiling existing code-bases to Rust. However, rule-based transpilation approaches are unidiomatic, fail to follow the target language conventions, and do not scale for larger programs. On the other hand, ML-based approaches (i.e., LLMs) cannot provide formal guarantees, thus removing the security benefits of Rust.
In this work, we show how to use both LLMs and formal verification to transpile verified and readable Rust programs. We evaluate our tool \tool by transpiling 1,394 programs from C++, C, and Go. Our tool with the Claude LLM can verify with bounded model checking for 40\% more programs for C++, 37\% for C, and 47\% for Go as compared to baseline.

\section{Data availability}
The software that supports the experiments, the benchmark programs, and the LLM Rust transpilations is available at \url{https://zenodo.org/records/10927704}. The TransCoder-IR model can be obtained from \url{https://github.com/facebookresearch/CodeGen}.

\bibliographystyle{ACM-Reference-Format}
\bibliography{reference}
\end{document}